\documentclass[12pt]{article}
\usepackage[authoryear]{natbib}
\usepackage{graphics, epsfig,colordvi,afterpage}
\usepackage{times}
\usepackage{amsmath}
\usepackage{mathabx}
\usepackage{paracol}

\input{epsf}

\def\be{\begin{equation}}
\def\ee{\end{equation}}
\def\bea{\begin{eqnarray}}
\def\eea{\end{eqnarray}}

\begin{document}
\baselineskip = 25 pt
\footskip = 0.75 in
\bibliographystyle{apalike}

\centerline{\bf Conduction velocity of intracortical axons in monkey primary visual cortex grows with distance:}
\centerline{\bf implications for computation}

\vskip 0.1 in
\centerline{Li Zhaoping, li.zhaoping@tuebingen.mpg.de }

\centerline{Max Planck Institute for Biological Cybernetics and University of T\"ubingen}

\vskip 0.1 in

\centerline{October 27, 2025}

\vskip 0.1 in

\section{Abstract}

A critical visual computation is to construct global scene properties from activities of 
early visual cortical neurons which have small receptive fields. 
Such a computation is enabled by contextual influences, through which a neuron’s response 
to visual inputs is influenced by contextual inputs outside its classical receptive fields.
Accordingly, neurons can signal global properties including visual saliencies and 
figure-ground relationships.  Many believe that intracortical axons conduct signals 
too slowly to bring the contextual information from receptive fields of other neurons. 
A popular opinion is that much of the contextual influences arise from feedback 
from higher visual areas whose neurons have larger receptive fields.  
This paper re-examines pre-existing data to reveal these unexpected findings:  
the conduction speed of V1 intracortical axons increases approximately linearly 
with the conduction distance, and is sufficiently high for conveying the contextual 
influences. Recognizing the importance of intracortical contribution to 
critical visual computations should enable fresh progress in answering long-standing questions.

\section{Introduction}

Receptive fields of primate V1 neurons are small.  In monkeys, their diameters are typically 
less than one degree in visual angle for parafoveal receptive fields, or less than 
two degrees for most V1 receptive fields \citep{HubelWiesel1968,GattassEtAl1987}.
However, a V1 neuron's  response to inputs within its receptive field can be 
influenced by contextual inputs which are up to several degrees away from this classical receptive 
field \citep{KnierimVanEssen1992, SceniakEtAl1999, AngelucciBressloff2006}. 
Such contextual influences enable V1 neurons to carry out important computations 
such as signaling visual saliency \citep{LiTICS2002}, which is a global 
scene property that depends on contextual inputs outside the receptive field. 

For example, a vertical bar is more salient in a contextual background of horizontal bars 
rather than vertical bars, because it  evokes a higher V1 neural response 
in the horizontal context than in the vertical context.  
This higher response in the horizontal context is due to iso-orientation suppression, 
so that the contextual or surround suppression on the neural response is stronger 
when the orientation of the contextual bars is close to the orientation of the bar 
within the receptive field \citep{KnierimVanEssen1992}.  
Indeed, when a monkey is searching for a uniquely oriented bar in a background of 
uniformly oriented bars, a faster onset of a saccade towards the target bar 
is typically preceded by a higher V1 neural response to this target bar \citep{YanZhaopingLi2018}.

Contextual influences are also observed in V2. For example, when a foreground figure 
surface occludes a background object surface, and when a V2 neuron’s receptive field 
covers a small segment of the occluding border between the foreground and background surfaces, 
this neuron’s response often  depends on whether the foreground  figure is at one or the 
other side of the border. In other words, this neuron’s response level can depend on 
contextual input information that is entirely outside its receptive field, and it 
conveys the information regarding which surface owns the border \citep{ZhouFriedmanHeydt2000}.
Such V2 signals for border ownership serve figure-ground computation for 
object segmentation \citep{VonderHeydtZhang2018}.

To enable the contextual influences, information about the contextual inputs outside the 
receptive field of a neuron must be transmitted to the neuron to impact its response. 
An important question is whether this transmission is through intracortical axons linking 
neurons with non-overlapping receptive fields or through top-down feedback axons from 
higher visual cortical areas. A neuron in a higher visual cortical area typically has 
a larger receptive field, which could cover visual locations inside and outside
the receptive field of a neuron in a lower visual cortical area. 
Hence a feedback axon from this higher cortical neuron to the lower cortical neuron could 
convey the contextual information.

For a neuron in V1, when visual inputs inside and outside its receptive field 
appear simultaneously,  the influence of the contextual inputs on its neural response starts 
at a latency of 10--20 milliseconds (ms) after the start of the of the response to the inputs 
within the classical receptive field \citep{KnierimVanEssen1992}. By comparing the latencies 
between influences from near versus far contextual inputs in monkeys, \citet{BairEtAl2003} estimated 
that the contextual influences effectively propagate across the distances on the cortical 
surface at a speed of around one meter/second (m/s). For V2 neurons, the latencies of the contextual 
influences appear insensitive to the distance of the context \citep{ZhouFriedmanHeydt2000,VonderHeydtZhang2018}. 

Meanwhile, previous works have concluded 
that the intracortical axons conduct at a speeds of 0.1--0.4 m/s, 
slower than the estimated speeds of around 3 m/s (or 2--6  m/s) by the 
inter-areal axons \citep{GrinvaldEtAl1994, BringuierEtAl1999, GirardEtAl2001, AngelucciBressloff2006}.
Consequently, many believe that conduction velocities of intracortical axons are too slow to 
bring contextual information in time. A popular opinion is that the contextual influences, particularly
influences from contexts further away from the receptive field,  
arise from feedback from higher visual areas \citep{VonderHeydtZhang2018, AngelucciBressloff2006}.

However, many previous estimates of the conduction speeds by intracortical axons are 
underestimates. Specifically, this conduction speed should be computed from the 
propagation latency,  i.e., the difference between the time of a spike at one location of an axon and the time
of this spike at another location of this axon as the result of the spike propagation along this axon.
Many \citep{GrinvaldEtAl1994, BringuierEtAl1999} have estimated this latency from latencies of 
neural responses to contextual visual inputs at various distances from the 
receptive fields \citep{NowakBullier1997}. However, these measured latencies 
include not only the time needed for signal propagation along the axons, 
but also the integration time, i.e., time  needed (typically several
milliseconds or longer) to integrate the inputs to charge up the membrane potentials of the 
post-synaptic neurons towards neural activation. This integration time applies not 
only to the post-synaptic neuron, but also be the neurons in the intervening neural 
circuit between the visual inputs and the recorded neural response.
Hence, this approach overestimates the latency, leading to an underestimation of the conduction speed. 

An accurate way to estimate the latency is by the latency of an antidromic 
spike relative to the time of an electrical stimulation at or near the axon of the 
same cell. The stimulation evokes a spike on the part of the axon very near 
(within $\sim 15 \mu m$) the tip of the stimulating 
electrode \citep{NowakBullier1998a, NowakBullier1998b, HistedEtAl2009, BakkumEtAl2013}, 
and this spike propagates along the axon and is recorded by the recording electrode somewhere downstream
from the stimulating electrode.
The technique of a collision test verifies that the recorded spike is the result of 
an antidromic propagation of the original, stimulation-evoked, spike towards the cell body 
without any intervening synapses \citep{NowakBullier1997}. 
However, this requires that the stimulating and the recording electrodes are very near 
the axon of the same neuron. This is easier to achieve for intercortical axons
between V1 and V2 by placing the two electrodes at retinotopically 
corresponding cortical locations. This is  very difficult to achieve for 
intracortical axons extending horizontally within V1, without adequate 
guidance for placing the two electrodes near an axon of the same cell.
\citet{GirardEtAl2001} was able to obtain the antidromic spike
across many pairs of electrodes between V1 and V2, but could only obtain
one such pair for intracortical axons.

For V1's horizontal axons, the best data so far for estimating the conduction speed 
are from the latencies by orthodromic propagation of action potentials recorded by \citet{GirardEtAl2001}.
Without a collision test, selecting recorded signals through the shape of the waveforms, 
and by considering only latencies that had a temporal jitter within the range $0.3$--$0.5$ ms \citep{GirardEtAl2001, GoldEtAl2006, BakkumEtAl2013}, 
the recorded spike was most likely post-synaptic to the axons of the stimulation-evoked spikes.
The range of the temporal jitter would be shorter if the recorded spike and the directly evoked spike 
by the electrical stimulation were on the same axon without an intervening synapse, and the jitter would be 
longer when a polysynaptic route is involved \citep{BullierHenry1980, BullierEtAl1988, NowakBullier1997, BakkumEtAl2013}. 
Meanwhile, the electrical stimulation,  a current pulse of 0.7 mA lasting for 0.2 millisecond 
(ms) \citep{GirardEtAl2001}, was many times the threshold level needed for evoking axonal spikes (the threshold is 
around $10\mu A$ or less at similar or shorter pulse durations \citep{ButovasSchwarz2003,HistedEtAl2009}). It 
should directly evoke axonal spikes in many neurons, such that numerous excitatory postsynaptic potentials (EPSPs) 
can converge nearly synchronously at the post-synaptic neuron near the recording electrode, 
and the time needed between the arrival of afferent spikes and the triggering of a spike 
in the post-synaptic neuron
could be only a fraction of a millisecond with a small jitter ($<0.5$ ms ) 
\citep{SingerEtAl1975,BullierHenry1979, NowakBullier1997, Xu-FriedmanRegehr2005}. 
In other words, using orthodromic spikes from artificial, supra-threshold electric stimulations, 
rather than neural responses to more naturistic visual input stimuli, the 
over-estimation of the latency by \citet{GirardEtAl2001}
could be limited to a fraction of a millisecond.

For each of the 156 pairs of  stimulating and recording sites in V1 of three monkeys,
 \citet{GirardEtAl2001} measured the latency $l$ of the orthodromic spike relative to the stimulation, 
the distance $d$ between the two electrodes, and obtained an estimation of the conduction speed (or velocity)
of the intracortical axons as $v=d/l$. They reported a median conduction 
speed of $0.33$ m/s using these orthodromic spikes. 
This median speed value for V1's intracortical conductions
is often quoted or used for analysis and arguments by previous works, 
for example, by \citet{AngelucciEtAl2002}, \citet{BairEtAl2003}, \citet{CarrascoEtAl2003}, \citet{AngelucciBressloff2006},
\citet{CraftEtAl2007}, \citet{JeheeEtAl2007}, \citet{VonDerHeydt2015}, \citet{MullerEtAl2018}, \citet{NurminenEtAl2018}, and \citet{FrankenReynolds2021}.   
However, \citet{GirardEtAl2001} did not analyze 
the relationship between the latency $l$ and the distance $d$,
and did not explicitly report the distances $d$ across their data sample. 
This paper re-analyzes these data kindly provided by Pascal Girard,
and highlights this relationship between the latency $l$ and the distance $d$
to discover the following unexpected finding: the conduction speed $v$ grows approximately 
linearly with the axonal conduction distance $d$. Accordingly, it further reveals that 
longer intracortical axons have their conduction speeds comparable to that in 
intercortical axons which transmit feedforward and feedback signals between 
different visual cortical areas.  Consequently, it argues that these intracortical axons are 
sufficiently fast to be a main player in mediating influences from near and far context.

\section{Methods}

\begin{figure}[tttthhhhh!!]
\centering
\includegraphics[width = 0.8\textwidth]{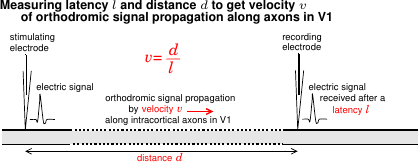}
\caption{\label{fig:Fig1}  The method to investigate the intracortical orthodromic conduction in V1 used
in \citet{GirardEtAl2001}. The conduction latency $l$ and distance $d$ were measured to obtain 
a conduction speed $v = d/l$.
}
\end{figure}

\citet{GirardEtAl2001} had 156 pairs of stimulating and recording sites in V1 of
three monkeys.  Electrical stimulation was by a 75-$\mu$m-tip tungsten microelectrodes assembled in a triple-
or double-electrode assembly. The tips delimited an equilateral triangle of $1.2$--$1.5$ mm side. They 
used cathodic current impulses, monophasic and unipolar, usually less than 1 mA (median, 0.7 mA) 
and with a duration of 0.2 ms.  The recording electrode was a tungsten microelectrode with a tip about 10 $\mu$m and recorded single unit spikes.  
The latency $l$ was defined as the time between the beginning of the electric 
stimulation artifact and the foot of the spike.  The jitter in $l$ was $0.3$--$0.5$ ms.
Using the waveform shape of the recorded spikes, they rejected spikes from passing-by axons so that
the recorded spikes were most likely from a neuron postsynaptic to the axons in which original 
spikes were evoked by the electrical stimulation.

The conduction speed for the orthodromic propagation along the V1 axon for each pair of electrodes 
is calculated as $v=d/l$ (Fig. 1).  \citet{GirardEtAl2001} found that the median and mean of the 
conduction speeds were $0.33$ m/s and $0.6$ m/s, respectively (Fig. 2B).
More details of the methods of the experiments were described in \citet{GirardEtAl2001}.

The matlab function  ``fit" was used for a linear fit between the latency $l$ and the distance $d$, and 
also for a linear fit between the velocity $v$  and the distance $d$. The slopes of the linear fits, 
as well as the 95\% confidence intervals of the slopes, were obtained.

\begin{figure}[tttthhhhh!!]
\centering
\includegraphics[width = 0.8\textwidth] {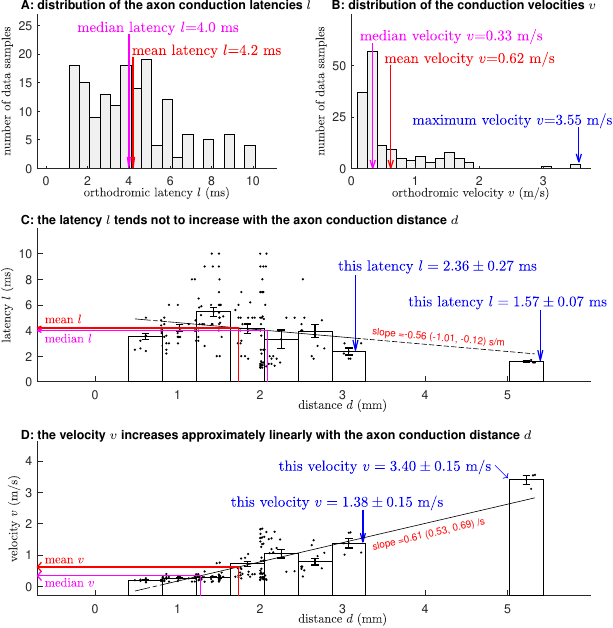}
\caption{\label{fig:Fig2}   Intracortical orthodromic conduction speed increases with axon length in V1 from the 
re-analysis of data from \citet{GirardEtAl2001}.  A and B: distributions of the 156 recordings of orthodromic propagation 
latencies $l$ and velocities $v$ between the stimulating and recording electrodes in V1.
C:   $l$ versus the distance $d$ between the stimulating and recording electrodes.  
Each dot is one of the 156 recordings.  Each bar is the mean $l$ across recordings in a range of $d$ 
specified by the borders of the bar.  Error bars are the standard errors of the means.   
The dashed line (with a slope and, in brackets, its 95\% confidence interval) is a linear fit of the data. 
D: like C, but for $v=d/l$.  C and D share the same data bins according to $d$. 
Across all data samples, the median and mean velocities are dominated by short $d$ samples. 
}
\end{figure}

\section{Results}

\subsection{The conduction speed in V1 axons increase with the propagation distance $d$}

Fig. 2A shows a histogram of the 156 latencies $l$. This histogram is the same as the bar 
histogram for horizontal V1 axons in Figure 3 of \citet{GirardEtAl2001} from their 
orthodromic activation data.  The median latency was $l=4.0$ ms (Fig. 2A).  
Our analysis revealed that the 156 samples were dominantly those with short 
inter-electrode distances  $d<2$ millimeters (mm).  Over-sampling of short $d$’s 
is unsurprising, since, given a stimulating electrode, a recording electrode at a 
shorter distance away is more likely to pick up the signal propagating down 
the axons. However, if the conduction speed $v$ depends on $d$,  then the average $v$  is 
biased by those of the short $d$ samples.  
Re-examining the data indeed found that $v$ increases about linearly with $d$, 
such that for $d> 5$  mm, the speed $v\ge 3$  m/s!

A linear fit of $l$ versus distance $d$ gives a negative slope $-0.56$ s/m (Fig. 2C). 
The 95\% confidence interval of this slope is $(-1.01, -0.12)$ s/m, entirely within 
the negative range (Fig. 2C).  If anything, the average latency $l$ decreased with 
distance $d$!  Among the three longest distance $d>5$  mm samples, the mean 
latency was $l=1.57\pm 0.07$ ms.  The mean latency for $d\approx 3$ mm samples 
was $l=2.36 \pm 0.27$ ms. A linear fit of conduction velocity $v$ versus $d$ 
(Fig. 2D) gives a positive slope $0.61$/s, with a 95\% confidence interval 
$(0.53,0.69)$/s, consistent with the idea that $v$ increases with $d$.  
Removal of the three $d>5$ mm samples gives the confidence interval $(0.4,0.6)$/s, 
which robustly remains in the positive range.  Indeed, for the  $d\approx 3$ mm 
samples, the velocity was $v=1.38 \pm 0.15$ m/s, four times as large as the 
median  $v=0.33$ m/s. Among our three $d>5$ mm samples, the average  $v=3.40 \pm 0.15$ m/s 
is similar to the speed of feedforward and feedback connections between 
V1 and V2 \citep{GirardEtAl2001}.  Among all the 156 $v$‘s (Fig.2BD), the median 
and mean $v$ were indeed dominated by the small $d<2$ mm samples.

\subsection{Considerations of possible factors contributing to the recorded data and the estimated speed}

We note that our estimated velocity $v$ may also be underestimates. If an axon has a 
trajectory with several changes of directions,  rather than straight, between the 
stimulating and recording electrodes, our distance $d$ would be shorter than the 
actual axon length, making our estimated $v$ smaller than the actual conduction velocity. 
Hence, the intracortical V1 axons could conduct at a faster 
speed than what we report here, this would only make our conclusion stronger.  

The recorded spike from a neuron is most likely the result of many
input spikes propagated by many different axons, and these input spikes converge almost 
synchronously onto the postsynaptic neuron being recorded.  The temporal jitter,  $0.3$-$0.5$ ms, 
in the latency of the recorded, post-synaptic, spike manifests the random (although nearly synchronous) 
temporal arrival times of the many input spikes \citep{Xu-FriedmanRegehr2005}.

We can ask whether it is likely that the recorded spikes involved a V1-V2-V1 route of
signal propagation from the electrical stimulation sites via the feedforward and feedback 
axons between V1 and V2.  For example, the electrical stimulation could evoke spikes in the feedforward axons 
towards V2, leading to postsynaptic V2 spikes which propagate orthodromically back to V1 before
activating V1 neurons whose spikes could then be recorded (since spikes from passing-by axons were 
rejected \citep{GirardEtAl2001}).
Another possibility could be that the electrically evoked spikes in V1 could propagate to V2 
antidromically through feedback axons, invading the soma of V2 neurons to cause 
somatic spikes, after an antidromic-to-orthodromic delay (of about $0.18$ ms, with a very small 
temporal jitter of $0.05$ ms \citep{SchmitzEtAl2001}), 
these spikes could then propagate orthodromically towards V1 via feedback fibers to 
activate post-synaptic V1 neurons.
Both possibilities should lead to a larger temporal jitter in the latency of the recorded 
V1 spike, compared to the temporal jitter if intracortical propagation was involved instead.
This is because the temporal jitter of the recorded neural spike could be sufficiently small 
only when the pre-synaptic inputs are sufficiently numerous and synchronous \citep{Xu-FriedmanRegehr2005}. 
Since V1 neurons receive about 10 times as many intracortical inputs as feedback inputs 
from V2 \citep{MarkovEtAl2011, SiuEtAl2021},  if the recorded spike were caused by signal 
propagation in feedforward and/or feedback axons (rather than intracortical axons), 
the number of pre-synaptic inputs would be much smaller so that a much larger temporal jitter would arise
in the recorded latency. 
Furthermore, when the V1-V2-V1 route involved an orthodromic rather than
an antidromic route to V2, this route would be disynaptic from the stimulation to the recorded 
spike, making the temporal jitter even larger than that ($0.3$--$0.5$ ms, consistent with a monosynaptic pathway) 
recorded in the data. This is because each synapse  gives additional temporal jitter through 
stochastic neurotransmitter releases. Hence this orthodromic V1-V2-V1 route is even less likely 
than the antidromic V1-V2-V1 route.  In any case, considering that the temporal jitter of the latency 
is $0.3$--$0.5$ ms in our data, it is quite unlikely that they involved feedforward and feedback axons 
between V1 and V2.

The V1-V2-V1 route would be even less likely for our data samples 
with distance $d>3$ mm,  since these data samples 
have short latencies $l\le 3$ ms, or even $l\approx 1.5$ ms for distance $d\ge 5$ mm (See Fig. 2C).  
Using antidromic spikes verified by a collision test,
\citet{GirardEtAl2001} showed that the latencies by  the intercortical conduction on 
the feedforward and feedback axons (without involving any synapses) between V1 and V2 
have a median of about $1.25$ ms (see Figure 1 of \citep{GirardEtAl2001} for a distribution of these latencies). 
Hence, the round-trip conduction latency, involving both the feedforward and feedback latencies, 
should have a median value around $2.5$ ms. 
In addition, the probability $p$ that the one-way latency was less 
than $1$ ms was $p\approx 0$ for the feedforward axons and $p\approx 0.04$ for the feedback axons, 
so that the round-trip latency has only a probability $p\approx 0.02$ to 
be $\le 1.5$ ms.  
Furthermore,  with $d > 3$ mm,  neurons near the stimulating electrode were unlikely
to have overlapping receptive fields with neurons
near the recording electrode. Since the feedforward and feedback connections are more
likely to link retinotopically corresponding neurons across different cortical areas, 
the number of pre-synaptic inputs from V2 to converge onto the recorded V1 neuron
via the V1-V2-V1 route should be further reduced, making it even less likely 
that the temporal jitter of the recorded latency to be as small as $0.3$--$0.5$ ms in our data.
Therefore, we conclude that the data samples with $d\ge 3 $ mm are much less likely
than the data samples with $d \le 2$ mm to involve the feedforward and feedback axons.

Among the data samples with $d \le 2$ mm, many have latencies $l>3 $ ms, making
it more difficult to rule out the possibility of a V1-V2-V1 route by the $l$ value 
alone for these high $l$ samples.  This possibility, although small by the argument of the 
temporal jitter above, implies that our estimated conduction speed of 
the intracortical fibers for our low $d$ samples may be quite inaccurate for unexpected reasons.  
In any case, our conclusion should still holds for our high $d$ data samples, for 
which intracortical connections conduct at speeds that are much faster than previously thought.

\section{Discussion}

When two V1 neurons are $d>2$ mm apart, they should be in different 
hypercolumns in V1 and most likely have non-overlapping receptive fields.  
In monkey V1, at eccentricity $E$, the area of the cortex
devoted to one unit area of visual field size is 
described by this  cortical magnification factor \citep{vanEssenEtAl1984}
\begin{equation}
M_a = 103 (0.82^o + E) ^{-2.28} \textrm{mm}^2/\textrm{degree}^2.
\end{equation} 
Accordingly, 
at eccentricities $E  = 5^o$, $10^o$,  $20^o$, and $40^o$, respectively,
a distance of $2$ mm (or $4$ mm)  in V1 spans roughly $1.5^o$, $3^o$, $6.3^o$, $13.5^o$  (or roughly 
$3^o$, $6^o$, $12.5^o$, $27^o$) in visual angle, 
while  the average size of the receptive fields is roughly $0.3^o$, $0.6^o$,  $1.1^o$, and $3.2^o$ 
in one dimension (according to Fig 7A of \citet{vanEssenEtAl1984}).
When the first neuron sends an intracortical axon to synapse on the second neuron, 
the activity of the second neuron can be influenced by the contextual visual inputs that are in the receptive field of the first neuron 
(and outside its own receptive field). Our findings suggest that intracortical connections linking these two neurons have 
high conduction speeds.  These speeds are sufficient for the fast propagation, at around $v=1$ m/s, of contextual influences 
observed physiologically \citep{BairEtAl2003}.  

It is likely that these long intracortical V1 connections are myelinated \citep{WaxmanBennett1972}.  
If one assumed a velocity $v$ for $d>2$ mm based on the median $v=0.33$ m/s dominated by $d<2$ mm samples, the latency for  
$d=3$ mm, $5$ mm, or  $10$ mm would be $9$ ms, $15$ ms, or $30$ ms, respectively. 
However, all our data samples have a latency  $l \le 10$  ms, 
sufficiently short for the physiologically observed contextual influences, which emerge 
about 10--20 ms after the start of the neural responses to visual inputs \citep{KnierimVanEssen1992}.  
It is therefore incorrect to assume that the intracortical connections are too slow for the 
contextual influences.

We found that conduction velocity of intracortical V1 axons grows approximately linearly with the conduction distance.   
Hence, when contextual influences are mediated by the intracortical axons, contextual inputs at different 
distances from a neuron can synchronize their influences on this neuron by a common latency.  
Since V1’s intracortical axons have a finite range of several millimeters, contextual inputs that are 
too far can exert their influences via the intervening contexts. For example, let A, B, and C
be three separate contour segments along a long and smooth object contour, such that the intracortical connections could 
directly link between segments A and B, and between segments B and C, but not between  segments A and C 
because the distance between A and C is longer than the longest possible intracortical axon.
Through mutually facilitative contextual influences,
the responses to contour segments A and B can directly enhance each other, so can the responses to segments 
B and C \citep{KapadiaEtAl1995, VonderHeydtZhang2018}. 
Through the intervening segment B, contour segments A and C can 
also enhance each other. Computational modeling of the neural circuits demonstrates that 
the mutual facilitation in such responses to contours emerges by a short latency as a collective
behavior of the neural circuit \citep{ZhaopingNeuron2005}. 
This short latency is not overly sensitive to the spatial extent of the contour, and is part of a 
collective phenomenon in a neural circuit with recurrent interactions that reinforce positive feedbacks between interacting elements.

Fast intracortical axons do not preclude top-down feedback from also contributing to the contextual 
computation \citep{BullierEtAl2001, AngelucciBressloff2006, GilbertLi2013, VonderHeydtZhang2018}.
However, they compel us to investigate the respective roles, and relative importance, of intracortical 
and feedback contributions. Both computational models and careful analysis of experimental data have 
suggested potential roles by both intracortical and feedback connections (e.g., 
 \citep{CraftEtAl2007, BushnellEtAl2011, LiangEtAl2017, ChavaneEtAl2022, DavisEtAl2024}). 

For example, some computational models \citep{LiTICS2002, ZhaopingNeuron2005} have demonstrated the feasibilities 
and potentials of the intracortical interactions for computing context-dependent signals for saliency 
and border-ownership in V1 and V2 neurons. Simultaneously recorded neurophysiological and behavioral 
data confirm that saliency signals for upcoming saccades emerge in V1 $40$ to $60$ ms from the appearance of 
visual inputs, around the time of initial peak neural responses \citep{YanZhaopingLi2018}.
These short latencies allow only the following as possible neural bases for saliency: 
(1) intracortical V1 mechanisms and (2) feedback only from visual areas that have response 
latencies shorter than $40$ to $60$ ms \citep{NowakBullier1997}. Previous data suggest that 
contextual influences in V1 do not depend on feedback from V2 for some stimuli \citep{HupeEtAl2001}, but, in 
responses to low-saliency inputs, are affected by feedback from MT \citep{BullierEtAl2001}.

In addition, when neural responses are at longer latencies or when context arises from 
animal’s task and internal state, feedback should contribute more 
significantly \citep{BullierEtAl2001, AngelucciBressloff2006, GilbertLi2013, 
YanZhaopingLi2018, VonderHeydtZhang2018}. For example, reversible inactivation of monkey MT showed 
that feedback connections serve to amplify and focus activity of neurons in lower-order areas 
for differentiating between figure and ground (Hupe et al 1998). Furthermore, while V2 neural 
responses can signal border ownership based on contextual inputs, memory-like persistence of 
this signal after the context is 
removed, and de novo transfer of this signal across saccades to responses of other 
neurons, suggest a strong role of feedback signals from higher visual areas such as 
V4 \citep{OHerronVonderHeydt2013, FrankenReynolds2025}. 
Meanwhile, it is preferable to refine the definition of ``contextual influences" and 
categorize them into different kinds. Some contexts are in the visual input image,  other contexts 
are from the animal's behavior and task, and the influence of a visual input context can be greatly 
modulated by attention and task \citep{FreemanEtAl2003, LiETAl2004}. Top-down feedback for visual 
recognition, using the computation of analysis-by-synthesis, are certainly 
present (see e.g., \citep{AlbersETAl2013, ZhaopingNewFramework2019, Zhaoping_iScience2025, XinEtAl2025}) 
and are likely to manifest in, e.g., figure-ground effects mediated by intracortical connections.
Understanding the roles of feedforward, lateral, and feedback connections in various kinds of
contextual influences is essential to understanding how vision works \citep{ZhaopingBook2014}.

In summary, this paper reports that the intracortical axons can be sufficiently fast, 
and their conduction speed increases with distances, allowing them  to convey information from contextual inputs nearly synchronously 
from near and far.  These findings should enable fresh progress  in answering long-standing questions on 
how neural circuits combine feedforward, horizontal, and feedback connections to carry out critical 
computations on global properties about visual scenes in the three-dimensional world based on local 
image features in two-dimensional retinal images. Furthermore, they should help us
dissect the roles of various circuit components for visual recognition and action modulated 
by attentional states and task demands.

\section{Acknowledgement}

I am very grateful to Pascal Girard for sharing the original data with me and for many discussions 
and communications via video meetings and emails, and for reading various versions of this manuscripts to provide feedback.  
This work is supported in part by the Max Planck Society and by the University of T\"ubingen.

\vskip 0.1 in


\end{document}